\documentclass[preprint,proceedings]{rmaa}

\SetYear{2002}
\SetConfTitle{The Eighth Texas-Mexico Conference on Astrophysics}

\title{Density Profiles of Dark Halos from their Mass Accretion Histories} 

\author{
  M. A. Alvarez,\altaffilmark{1,2} 
  K. Ahn,\altaffilmark{1}
  and P. R. Shapiro\altaffilmark{1}}

\altaffiltext{1}{Department of Astronomy, University of Texas at Austin,
        Austin, TX 78712}
\altaffiltext{2}{DOE Computational Science Graduate Fellow}

\shortauthor{Alvarez, Ahn, \& Shapiro}
\shorttitle{Dark Halo Density Profiles}

\fulladdresses{
\item Marcelo A. Alvarez, Kyungjin Ahn, Paul R. Shapiro: 
        Department of Astronomy, University of Texas at Austin, 
        Austin, TX 78712, USA (marcelo,kjahn,shapiro@astro.as.utexas.edu).}
\suppressfulladdresses
\listofauthors{M. A. Alvarez, K. Ahn, \& P. R. Shapiro}
\indexauthor{Alvarez, M. A.}
\indexauthor{Ahn, K.}
\indexauthor{Shapiro, P. R.}

\abstract{
We use the universal mass accretion history recently reported for
simulations of halo formation in the cold dark matter model (CDM) to
analyze the formation and growth of a single halo.
We derive the time-dependent density profile three different ways, based
upon three approximations of successively greater realism: equilibrium,
radial orbits, and a fluid approximation.  For the equilibrium
model, the density profile is well-fit by either an NFW or Moore profile
over a limited range of radii and scale factors.  For the radial orbit
model, we find profiles which are generally steeper than the NFW profile,
with an inner logarithmic slope approaching -2, consistent with a purely
radial collisionless system.  In the fluid approximation, we find
good agreement with the NFW and Moore profiles for radii resolved by N-body
simulations ($r/r_{200}\geq 0.01$), and an evolution of
concentration parameter nearly identical to that found in N-body
simulations.  The evolving structure of cosmological halos is therefore 
best understood as the effect of a time-varying rate of mass infall on a 
smoothly distributed, isotropic, collisionless fluid.}

\addkeyword{Cosmology: Theory} \addkeyword{Dark Matter}
\addkeyword{Galaxies: Formation} \addkeyword{Galaxies: Kinematics and
Dynamics} \addkeyword{Large-scale Structure of Universe}

\begin{document}
% Typeset article header
\maketitle

\section{Introduction}
Current understanding of dark matter halos relies upon N-body simulations of
collisionless cold dark matter (CDM) with Gaussian-random-noise initial 
conditions.  Two ``universal'' profiles bracket the results ; (Navarro, Frenk, 
\& White 1997; Moore et al. 1998).  The NFW(Moore) profile has an inner 
density profile $\rho\propto r^{-1}(r^{-1.5})$.
Wechsler et al. (2002) (WBPKD hereafter) found that the mass and concentration
of individual N-body CDM halos grow over time according to simple universal 
formulae.  In what follows, we attempt to explain this result. 

\begin{figure}[!t]
  \begin{center}
  \includegraphics[width=3.3in]{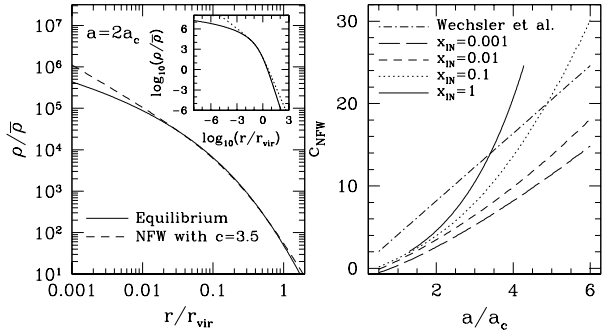}
  \end{center}
  \caption{(Left) Density profile from equilibrium model along with 
best-fitting NFW profile for this profile at present.  Inset in upper-right 
shows same over much larger range. (Right) Evolution of NFW concentration 
parameter in the equilibrium model. Different line types indicate different 
ranges $x_{\rm in}<x<1$, within which halo was fit to an NFW profile, 
where $x\equiv r/r_{\rm vir}$,$r_{\rm vir}\equiv r_{200}$.
}
  \label{fig:fig1}
\end{figure}
\begin{figure}[!b]
  \begin{center}
  \includegraphics[width=2.7in]{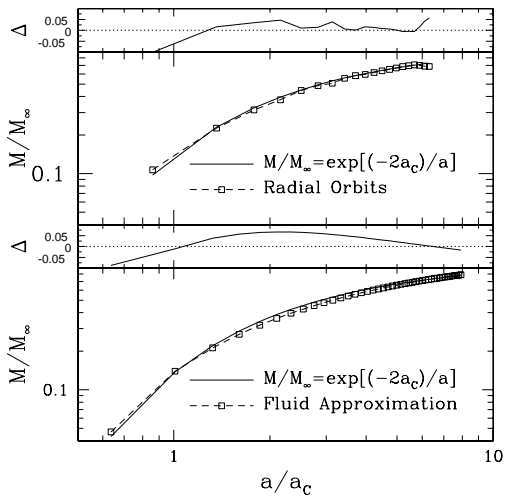}
  \end{center}
  \caption{Evolution of mass for the radial orbits (top) and fluid 
approximation (bottom) simulations.  Shown above each are the fractional 
deviations $\Delta \equiv (M_{\rm exact}-M)/M$.
}
  \label{fig:fig1}
\end{figure}
\begin{figure}[!t]
  \begin{center}
  \includegraphics[width=2.5in]{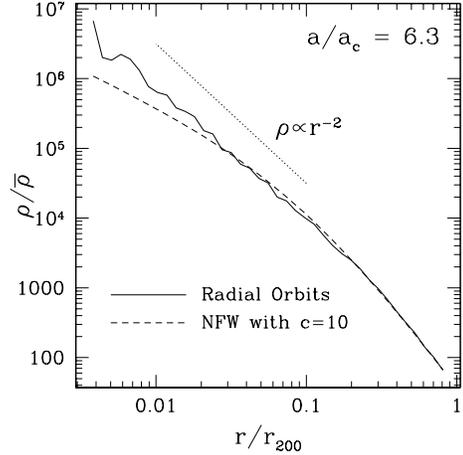}
  \end{center}
  \caption{Density profile at the end of the radial orbit simulation.}
  \label{fig:fig3}
\end{figure}

\section{Three Different Models}
We attempt to understand the form and evolution of dark matter
halos 
with three spherically-symmetric 
models; equilibrium, radial orbits, and an isotropic 
fluid. Each model assumes that the
mass $M_{\rm vir}$ within an overdensity $\delta_{\rm vir}$ follows
the relation given by WBPKD
\begin{equation} 
\label{wec}
M_{\rm vir}(a)=M_\infty\exp\left[-Sa_{\rm c}/a\right], 
\end{equation}
where $a_c$ is the scale factor at collapse and $S$ is the logarithmic mass 
accretion rate $d({\rm ln}M_{\rm vir})/d({\rm ln}a)$ when $a=a_c$. 
Such a relation is claimed to
be a good fit to the evolution of halos of different masses and formation
epochs.  As in WBPKD, we have chosen $S=2$ and 
$\delta_{\rm vir}=200$, so that the halo has a mass 
$M_{\rm 200}$ and radius $r_{200}$.
We have found an initial perturbation profile consistent with equation 
(1),
\begin{equation}
\frac{\delta M}{M}\equiv\frac{M-\overline{M}}{\overline{M}}=\delta_{\rm i}
{\rm ln}\left[\frac{M}{bM_\infty}\right],
\label{init}
\end{equation}
where $\delta_{\rm i}$
depends on the initial scale factor $a_{\rm i}$ and $\delta_{\rm vir}$, and
$\overline{M}$ is the unperturbed mass.  The free parameter $b$ is unity in the
absence of pressure or shell crossing inside of $r_{\rm vir}$.

\subsection{Equilibrium} 
In the simplest model, we have made the assumption of complete
equilibrium inside the halo, so that the velocity is zero for $r<r_{\rm
vir}\equiv r_{200}$.  The mass of the halo is given by
\begin{equation} 
M_{\rm vir}(a)=\frac{4\pi}{3}\delta_{\rm vir}\overline{\rho}r_{\rm vir}^3, 
\label{vmass}
\end{equation} 
where $\overline{\rho}$ is the mean density at that epoch.  
Mass continuity implies the density $\rho_{\rm vir}$ just inside the virial 
radius is
\begin{equation}
\label{jump}
\frac{dM_{\rm vir}}{da}=4\pi\rho_{\rm vir}r_{\rm vir}^2\frac{dr_{\rm vir}}{da}.
\end{equation}
Differentiating equation (3) and combining with equations (1) and (4) 
yields
\begin{equation}
\label{rhovir}
\frac{\rho_{\rm vir}}{\rho_0}=\delta_{\rm
vir}a^{-3}\left[1+\frac{3a}{Sa_c}\right]^{-1},
\end{equation} 
where $\rho_0$ is the mean background density at $a=1$.  The virial
radius is given by
\begin{equation}
\label{rvir}
\frac{r_{\rm vir}}{r_0}=a {\rm
exp}\left[\frac{-Sa_c}{3}\left(\frac{1}{a}-1\right)\right].
\end{equation}
Equations (5) and (6) are parametric in $a$, implying
a radial density profile $\rho(r)=\rho_{\rm vir}(r_{\rm vir})$ which is
frozen in place as matter crosses $r_{\rm vir}$.
Taking the limit in which $a\rightarrow\infty$, the outer density profile 
approaches $\rho\propto r^{-4}$ at late times, consistent with finite mass, 
while the inner slope becomes asymptotically flat.  
The NFW profile is given by
\begin{equation}
\label{nfw}
\frac{\rho(x)}{\overline{\rho}}=\frac{\delta_{\rm vir}g(c)}{3x(1+cx)^2},
\end{equation}
where
\begin{equation}
g(c)=\frac{c^2}{{\rm ln}(1+c)-c/(1+c)},
\end{equation}
and $x\equiv r/r_{\rm vir}$.  Combining equations (5) and
(7) with $x=1$, yields an equation for the
evolution of concentration with scale factor (see Fig. 1),
\begin{equation}
\label{cofa}
\frac{a}{a_c}=S\left[\frac{(1+c)^2}{g(c)}-\frac{1}{3}\right].
\end{equation}

\begin{figure}[!t]
  \begin{center}
  \includegraphics[width=2.7in]{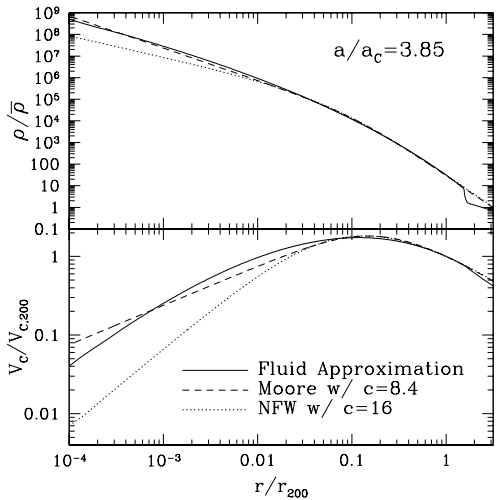}
  \end{center}
  \caption{(Top) Density profile at the end of the isotropic fluid 
	calculation. (Bottom) Circular velocity profile.}
  \label{fig:fig4}
\end{figure}
\subsection{Radial Orbits}
We use a finite-difference spherical mass shell code to follow the
evolution of a small amplitude perturbation which is chosen so that the
resulting virial mass will evolve according to equation (1) from an 
initial perturbation given by equation (2). 
The shell code has an inner reflecting core and the results presented here
used 20,000 shells (see Figs. 2 and 3).

\subsection{Fluid Approximation} 
The collisionless Boltzmann equation in spherical symmetry yields fluid 
conservation equations ($\gamma=5/3$) when random motions are isotropic.
Halos in N-body simulations have radially biased 
random motion, but the bias is small, especially in the center. This model is 
therefore a better approximation to halo formation in N-body simulations than 
one with purely radial motion. We use a 1-D, spherical, Lagrangian 
hydrodynamics code as in Thoul \& Weinberg (1995), using 1,000 zones 
logarithmically spaced in mass (see Figs. 2, 4, and 5).  The initial 
conditions were chosen in the same way as those for the radial orbit model 
(Eq. 2), with zero initial temperature.

\section{Results} 
\begin{itemize}
\item Equilibrium model does not reproduce linear evolution of 
concentration parameter with scale factor reported by WBPKD,
but can be fit by an NFW profile over a limited range of radii and scale 
factors (Fig. 1).  
\item Mass evolution for a perturbation given by equation 
(2) is close to that of equation (1) (Fig. 2).
\item The radial orbit model fails to reproduce the inner slope of the NFW
profile, approaching $\rho\propto r^{-2}$ instead, consistent with the argument
of Richstone \& Tremaine (1984) (Fig. 3).
\item Fluid approximation halo is well-fitted by NFW and Moore profiles for 
radii resolved by N-body simulations ($r/r_{200}\geq 0.01$) 
(Fig. 4).
\item Evolution of NFW concentration parameter in the fluid approximation is
a close match to that of WBPKD, with $c_{\rm NFW}=4.25a/a_c$
a good fit (Fig. 5). WBPKD reported $c_{\rm NFW}=4.1a/a_c$.
Complicated merging process is not necessary in order to understand density
profile evolution.
\end{itemize}
\begin{figure}[!t]
  \begin{center}
  \includegraphics[width=2.7in]{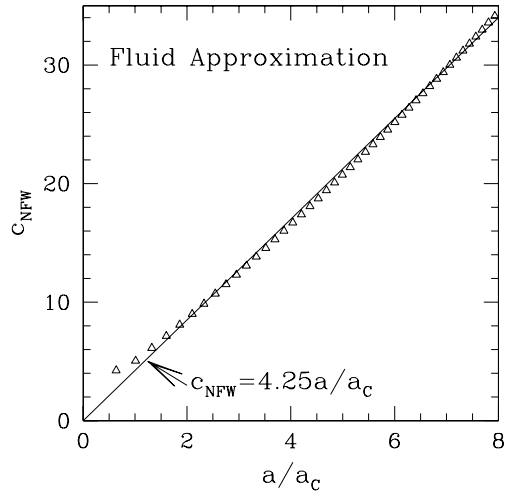}
  \end{center}
  \caption{Evolution of concentration parameter with scale factor in the 
isotropic fluid calculation.}
  \label{fig:fig5}
\end{figure}
\acknowledgments This work was supported in part by
grants NASA ATP NAG5-10825 and NAG5-10826 and Texas Advanced Research
Program 3658-0624-1999.  

\end{document}